# When Streams of Optofluidics Meet the Sea of Life

*Luke P. Lee*
*University of California, Berkeley / National University of Singapore*

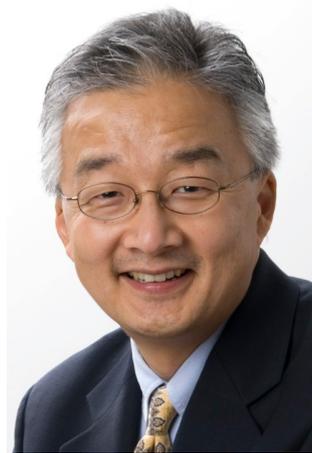

*Prof. Luke P. Lee received both his BA and PhD from UC Berkeley. He joined the faculty at the UC Berkeley in 1999 after more than a decade of industry experience. He became the Lester John and Lynne Dewar Lloyd Distinguished Professor of Bioengineering in 2005. He also served as the Chair Professor in Systems Nanobiology at the ETH Zürich from 2006 to 2007. He became Arnold and Barbara Silverman Distinguished Professor at Berkeley in 2010 and was reappointed again 2015. Lee is now a Tan Chin Tuan Centennial Professor, Director of the Biomedical Institute for Global Healthcare Research & Technology (BIGHEART), and Associate President (International Research and Innovation) at the National University of Singapore. Lee is a Fellow of the Royal Society of Chemistry and the American Institute of Medical and Biological Engineering. His work at the interface of biological, physical, and engineering sciences for medicine has been recognized by many honors including the IEEE William J. Morlock Award, NSF Career Award, Fulbright Scholar Award, and the HoAm Prize. Lee has over 350 peer-reviewed publications and over 60 international patents filed. His current research interests are quantum biological electron transfers in living organisms, molecular diagnostics of neurodegenerative diseases, and in vitro neurogenesis, with a focus both on studying fundamental quantum nanobiology and on solving ill-defined problems of global healthcare.*

## Abstract

In the vast sea of life sciences, the emergence of microfluidics and plasmon optics, like intrepid explorers, has enabled us to venture into uncharted territories. These fresh domains offer unprecedented possibilities in generating new quantitative biological understanding and precision medicine paradigms. Microfluidics and plasmon optics approaches empower detailed investigation at the molecular, sub-cellular, single-cell, and tissue network level. They also provide high-content molecular-level resolution and detail that is crucial for the advancement of precision medicine. These new insights and innovative inventions that offer superior solutions for life sciences research and precision healthcare have been made possible through the convergence of knowledge in biology, physical sciences, engineering, and medicine in the following areas:

## Single Cell Analysis

Pioneering work in microfluidics has enabled us to make new discoveries in the sea of cellular heterogeneity. Discerning cellular heterogeneity is important for understanding fundamental cellular processes and translational medicine research in the areas of aging, stem cells, and cancer. To overcome the technical limitations associated with single-cell culture and measurements, dynamic single cell analysis platforms were created by our group to distinguish cellular heterogeneity [1-5].

A higher resolution two-stage system to define cellular heterogeneity was also recently described by us [6]. In the first stage, multiplexed single-cell RNA cytometry was performed in microwell arrays containing over 60,000 reaction chambers. In the second stage, RNA cytometry data was used to determine cellular heterogeneity by defining a likelihood score. Monte-Carlo simulation and RNA cytometry data enabled calculation of the minimum number of cells required for detecting cellular heterogeneity. This experimental paradigm was applied to characterize RNA distributions of ageing-related genes in a highly purified mouse haematopoietic stem cell population. A novel group of genes that contributed to cellular heterogeneity was identified with this approach. It was also established that changes in the expression of certain genes like Birc6 during cellular ageing were attributable to a shift in relative cell proportions in high-expressing subgroups versus low-expressing subgroups. Microfluidic single cell analysis platforms have prompted new approaches for studies in ageing, stem cell biology, and cancer cell biology.

Concurrent transcript and protein quantification is conceptually appealing because of the potential to elucidate cryptic properties of biological systems that are not accurately represented by either mRNA or protein analysis alone. For applications in cancer, a microwell-based cytometric method was developed for simultaneous measurements of gene and protein expression dynamics in thousands of single cells [7]. The regulatory effects of transcriptional and translational inhibitors of cMET mRNA and protein in cell populations were quantified. Next, the dynamic responses of individual cells to drug treatments were studied by measuring cMET overexpression in individual non-small cell lung cancer (NSCLC) cells with induced drug resistance. Distinct correlated transcript– protein signatures emerged across NSCLC cell lines with a specified protein expression profile. This platform is ideal for interrogating the dynamics of gene expression, protein expression, and translational kinetics at the single-cell level, and represents a paradigm shift towards discovering vital cell regulatory mechanisms.

## Microphysiological Analysis Platforms (MAPs)

Integrated microfluidics systems are a key component of MAPs that provide clear navigation through the turbulent waters of drug discovery. MAPs are physiologically relevant dynamic 3D cell culture microfluidic systems and artificial organ models on-chip [8-11]. The first artificial liver sinusoid model on-chip described was instrumental in starting a scientific movement immersed in developing organ on-chip platforms [11]. Improved patient and disease-specific iPSC- based organ on-chip platforms for personalized medicine research are currently being developed [12].

Drug discovery and development are hampered by high failure rates attributed to an overreliance on non-human animal models for safety and efficacy testing. A fundamental problem inherent in this approach is that non-human animal models are not adequately representative of human biology. Consequently, there is an urgent need for high-content in vitro systems that accurately recapitulate physiological microenvironments and predict drug-induced toxicity reliably. Organ systems, which predict cardiotoxicity, are of principal importance because approximately one third of safety-based pharmaceutical withdrawals are due to cardiotoxicity issues.

A cardiac MAP with the characteristics of an ideal in vitro system for predicting cardiotoxicity was recently described [12]. Salient features of this cardiac MAP platform include cells with a human genetic background, physiologically relevant tissue architecture (e.g. aligned cells), computationally predictable perfusion mimicking human vasculature, and multiple modes of analysis (e.g. biological, electrophysiological, and physiological). The cardiac MAP can keep human induced pluripotent stem cell derived cardiac tissue viable and functional over extended periods. Pharmacological studies using the cardiac MAP platform showed half maximal inhibitory/effective concentration values (IC50/EC50) that are consistent with tissue and cellular scale reference data. MAPs are being adopted for human disease modeling, drug development, and toxicology. Currently, mini-brain MAP and pancreatic islet MAP devices are also being researched.

## Optofluidics for Application Specific Integrated Systems (OASIS)

Explorations in the realm of light-controlled microfluidics or optofluidics has led to the creation of sensitive and specific OASIS technologies for molecular diagnostics. The rapid diagnosis of infectious and human diseases requires an integrated molecular diagnostic platform with specific sample preparation units coupled to fast and sensitive detection methods for DNA, RNA, and protein biomarkers. Several, stand-alone, self-powered, and integrated microfluidic blood analysis platforms were established for the advancement of low-cost molecular diagnostic healthcare systems [13-19]. A handheld point-of-care genomic diagnostic system fulfilling the requirements and infrastructural limitations in resource poor settings and for use both developing and developed countries was recently described by us [15].

In the domain of label-free detection, integrated optofluidics has helped advance single molecular level surface enhanced Raman spectroscopy and nanoplasmonic PCR technologies. Improvements in PCR nucleic acid amplification and quantification are important because PCR remains the most sensitive, powerful, and widely used diagnostic tool in existence. In order to address problems associated with conventional PCR thermal cyclers, such as, fast cycling speed, uniform heating and cooling, and large heating block mass, an innovative optofluidic cavity PCR method using light- emitting diodes was developed [20, 21]. Ultrafast photonic PCR technology is based on plasmonic photothermal light-to-heat conversion via photon–electron–phonon coupling. An efficient photonic heat converter was designed and constructed using a thin Au film due to plasmon-assisted high optical absorption features of Au film. The plasmon-excited Au film is capable of rapidly heating the surrounding solution. Using this method, ultrafast thermal cycling (30 cycles) from 55 oC for annealing to 95 oC for denaturation can be accomplished within 5 min. This simple, robust and low cost approach to ultrafast PCR with efficient photonic-based heating can be readily integrated into a variety of devices or methods, including on-chip thermal lysis for sample preparation and heating for isothermal amplification. This optofluidic cavity PCR is highly desirable for rapid and accurate molecular diagnostics and precision medicine applications. An integrated OASIS platform with different sample preparation units will contribute to transformative life sciences experimental methodologies and molecular diagnostics. These concepts will be especially important for the development of lightweight, point-of- care devices for use in both developing and developed countries.

The life sciences are a vast sea, yet optofluidics has guided and empowered us to explore previously uncharted waters. This knowledge base has become a widely sought-after companion when venturing into new realms of science and technology. New holistic systems and improved adaptive and scalable experimental strategies for precision medicine can be derived from optofluidics while inspirations from biology can guide us in the design and development of these technologies [22-25]. Collectively, these strategies will provide unparalleled ability to perform experiments which were once ethically and technically unfeasible. Overall, the fundamental knowledge gained from this research can be readily translated into new technologies, theranostics, treatment paradigms, and personalized medicine solutions to improve the quality of life globally.

## References

1. Lab on Chip (2006) 6: 1445-1449
2. Analytical Chemistry (2006) 78: 7918-7925
3. Analytical Chemistry (2006) 78: 4925-4930
4. Lab on Chip (2007) 7: 457-462
5. PNAS (2012) 109: 4916-4920
6. Nature Communication (2014) 5: 3451
7. Lab on Chip (2016) 16: 3682
8. Lab on a Chip (2005) 5: 44-48
9. Biotechnology & Bioengineering (2005) 89: 1-8
10. Biotechnology & Bioengineering (2006) 94: 5-14
11. Biotechnology & Bioengineering (2007) 97: 1340-1346
12. Scientific Reports (2015) 5, Article number: 8883
13. Lab Chip (2008) 8: 2071-2078
14. Nature Nanotechnology (2009) 4: 742-746
15. PLOS One (2013) 8: e70266
16. ACS Nano (2012) 6: 7607-7614
17. ACS Nano (2013) 7: 6268-6277
18. Nature Communications (2013) 4: 2182
19. Lab Chip (2014) 14: 2287-2292
20. Light: Science & Applications (2015) 4: e280
21. Advanced Healthcare Materials (2016) 5: 167-174
22. Science (2005) 310: 1148-1150
23. Nature Materials (2006) 5: 27-32
24. Science (2006) 312: 557-561
25. Science (2013) 341:247-248